# A COMPLEXITY MEASURE BASED METHOD FOR STUDYING THE DEPENDANCE OF $^{222}$RN CONCENTRATION TIME SERIES ON INDOOR AIR TEMPERATURE AND HUMIDITY

DT Mihailovic[1,*], V Udovičić[2], M Krmar[3] and I Arsenić[1]

[1]Faculty of Agriculture, University of Novi Sad, Dositeja Obradovica Sq. 8, 21000 Novi Sad, Serbia
[2]Institute of Physics, University of Belgrade, Belgrade, Serbia
[3]Faculty of Sciences, Department of Physics, University of Novi Sad, Dositeja Obradovica Sq. 5, 21000 Novi Sad, Serbia

[*]E-mail: guto@polj.uns.ac.rs                    Telephone:+381216350552

**Abstract**

We have suggested a complexity measure based method for studying the dependence of measured $^{222}$Rn concentration time series on indoor air temperature and humidity. This method is based on the Kolmogorov complexity (KL). We have introduced (i) the sequence of the KL, (ii) the Kolmogorov complexity highest value in the sequence (KLM) and (iii) the KL of the product of time series. The noticed loss of the KLM complexity of $^{222}$Rn concentration time series can be attributed to the indoor air humidity that keeps the radon daughters in air.

## 1. Introduction

Radon ($^{222}$Rn) is the decay product of radium ($^{226}$Ra), and both elements are members of the uranium series ($^{228}$U). After generation from the radioactive decay of $^{226}$Ra, mostly in the earth crust, can be transported to the large distances and accumulated indoor due to the fact that radon is noble gas having no affinity to chemical reactions and relative long half life of 3.82 days. It is believed that, after smoking, radon is the next significant source of the lung cancer. The soil and the building materials are the most important source of indoor radon in dwellings. Indoor radon concentrations exceeding the level prescribed can cause possible health hazard of dwelling people (Jelle et al., 2011). Besides in the dwelling control and possible reduction of $^{222}$Rn presence is crucial in the area of low-background laboratories. Namely, radioactive gas radon, with its progenies, which emanates from the soil and construction materials, contributes significantly to the background radiation. In number of experiments where some measurable effects of low-probability process were followed, reduction of background radiation is often most significant way to improve sensitivity (Antanasijević et al., 1999; Dragić et al., 2011; Udovičić et al., 2009; Garcia et al., 1998; Jovancevic and Krmar, 2011).

Measured value of $^{222}$Rn concentration in some room is a final outcome of plenty processes including its generation, transport, accumulation and decay. Therefore, sometimes it is not possible to follow influence of different parameters, through the aforementioned processes, on the final result of radon measurement. In the recent time this problem is considered through: (i) studies of the dynamic of radon changes in some room with idea to find correlation with measurable parameters, mostly indoor environmental ones and (ii) development of models for predicting the radon concentration and dynamics in some dwellings. Let us note that those designed models, even in the cases when affecting processes are treated on a simplified manner, operate with large number of quantities (Collignan et al., 2012; Girault and Perrier, 2012).

Analysis of $^{222}$Rn concentration time series is an important step in deriving conclusions about interaction radon – environment. These analyses are mostly based on relatively simple statistical methods, which in some segments give a clear picture about this interaction. However, if we claim to get more insight we have to apply the comprehensive mathematical procedures in analysis of those time series. Thus, illustrative examples for that kind of approach are papers by Negarestani et al. (2003) and Seftelis et al. (2008). Negarestani et al. (2003) have proposed a new method based on adaptive linear neuron to estimate the radon concentration in soil associated with the environmental parameters. Seftelis et al. (2008) have developed a mathematical function to describe the diurnal variation of radon progeny. Our intention is to offer a a complexity measure based method for establishing the dependance of $^{222}$Rn concentration time series on indoor environmental parameters. A possible field of application of this method is not restricted only on either indoor or outdoor radon time series. Moreover, this mathematical procedure is applicable in analysis of time series, obtained by the measurements, for which we should establish whether influence of some parameter makes the distribution of measured quantity less or more stochastic.

In this paper we consider the dependance of $^{222}$Rn concentration on indoor parameters, in particular on air temperature and humidity, through dynamics of a complex system, which can be analyzed from the signal, sent in the form of time series of measured values. In that sense we develop a complexity measure based method that help us to get an insight into the complexity of the $^{222}$Rn concentration time series in dependance on indoor air temperature and humidity that can not be done by traditional mathematical statistics. Kolmogorov Complexity (KL in further text) is used in order to describe the complexity or degree of randomness of a binary string. It is in the literature also known as stochastic complexity or descriptive complexity. This measure was developed by Andrey N. Kolmogorov in the late 1960s (Li and Vitanyi, 1997). Later following Kolmogorov's idea, Lempel and Ziv (1976) developed an algorithm for calculating the measure of complexity. We will refer to the Lempel-Ziv Algorithm by LZA. It can be considered as a measure of the degree of disorder or irregularity in a time series. The LZA measure has been used for evaluation of the randomness present in time series. Entropy is commonly used to characterize the complexity of a different time series also including $^{222}$Rn concentration ones. However, to our knowledge, the KL measure has not been used for analyzing the complexity of $^{222}$Rn concentration time series. In this study we also use the Sample Entropy (SE in further text) proposed by Richman and Moorman (2000), which is unbiased and less dependent on data although traditional entropies quantify only the regularity of time series having some disadvantages (Chou, 2012).

The purpose of this paper is to introduce a complexity measure based method for studying the dependance of $^{222}$Rn concentration time series on indoor parameters (in particular, air temperature and humidity). For that purpose we have used indoor $^{222}$Rn concentration time series measured during the 2009 in the Low-Background Laboratory for Nuclear Physics at the Institute of Physics in Belgrade. Comparing complexities of $^{222}$Rn concentration (Rn), indoor air temperature (T) and humidity (H) time series we study the dependence of Rn on T and H. It is done through the following steps. In Section 2: (i) we describe the KL, introducing the calculation of its sequence for a time series (2.1), (ii) we introduce Kolmogorov complexity highest value in the sequence (KLM) in a time series. In Section 3 we describe the laboratory and experimental details. In Section 4 we show results, which include statistical evaluation and discussion. Concluding remarks and some recommendation are given in Section 5.

# 2. The Kolmogorov complexity measure based method for establishing the level of dependance of $^{222}$Rn concentration time series on environmental indoor parameters

## *2.1. Kolmogorov complexity*

Kolmogorov complexity (KL in the further text) is a measure, which points out to the minimum length of a program such that a universal computer can generate a specific sequence. A good introduction to the KL can be found in Ref. (Cover and Thomas, 1991) and with a comprehensive description in Ref. (Li and Vitanyi, 1997). On the basis of Kolmogorov's idea, Lempel and Ziv (1976) developed an algorithm (LZA), which is used in assessing the randomness of finite sequences as a measure of its disorder. Let us note that the KL complexity is not able to distinguish between time series, which have different amplitude oscillations but very similar random components. Recently, the KL is intensively used in analysis of different kind of biomedical, hydrological and environmental time series. Namely, the complexity of these time series may be lost due to the different reasons that come from reducing the functionality of system segments represented by those time series. For example, Gomez and Hornero (2010) using entropy and complexity analyses of Alzheimer's disease have showed that the complexity reduction seems to be associated with the deficiencies in information processing suffered by AD patients. Another example, the river flow time series analysis by Orr and Carling (2006) point out that the complexity loss may be attributed to the extent of human intervention involving land, urbanization, commercial navigation and other activity.

## 2.2. The Kolmogorov complexity sequence

The computation of the KL we perform is slightly different than it is usual in this kind of analysis. Therefore, environmental signal in the form of time series we transformed into a finite symbol string by comparison with series of thresholds $\{x_{t,i}\}$, $i = 1, 2, 3, 4, ..., N$, where each sample is equal to the corresponding sample in the considered time series $\{x_i\}$, $i = 1, 2, 3, 4, ..., N$. The original time series samples are converted into a 0-1 sequences $\{S_i^{(k)}\}$, $i = 1, 2, 3, 4, ..., N$, $k = 1, 2, 3, 4, ..., N$ defined by comparison with a threshold $x_{t,k}$,

$$S_i^{(k)} = \begin{cases} 0 & x_i < x_{t,k} \\ 1 & x_i \geq x_{t,k} \end{cases} \qquad (1)$$

After we apply the LZA on each sample of series $\{S_i^{(k)}\}$ we get a sequence of the KL complexity $\{K_i^C\}$, $i = 1, 2, 3, 4, ..., N$. This sequence we introduce to explore the range of amplitudes in a time series representing a process, for which that process has the highest complexity. In our case it is the $^{222}$Rn concentration. The highest value $K_m^C$ in this sequence, i.e. $K_m^C = \max\{K_i^C\}$, we call the Kolmogorov complexity highest value in the sequence (KLM).

## 2.3. Kolmogorov complexity of the product of two time series

As it mentioned above, the complexity of a time series can be lost due to reduction in functioning the system or process represented by that time series. It means that there exists a source, which causes that time series becomes rather uniform than random. For example, let us suppose that one physical process (in our case that is detection of indoor $^{222}$Rn concentration) is under influence of some parameters (in our case they are indoor air temperature and humidity). Which of these parameters contributes to reducing the complexity of detection indoor $^{222}$Rn concentration, we can establish through computing the complexity of the product of two or more time series obtained by measurements.

Let us suppose that we have two independent and positive definite time series $\{x_i\}$ and $\{y_i\}$, $i = 1, 2, 3, 4, ..., N$, which are generated either computationally or by a measuring procedure. We define product $\{z_i\}$ of two $\{x_i\}$ and $\{y_i\}$ as $\{z_i\} = \{x_i y_i\} = (x_1 y_1, x_2 y_2, x_3 y_3, ..., x_N y_N)$, where amplitudes in each time series are normalized on its highest value and thus taking the values in interval $(0,1)$. Further, with $K^C(x_i)$, $K^C(y_i)$ and $K^C(z_i)$ we define the KL complexities for corresponding time series $\{x_i\}$, $\{y_i\}$ and $\{z_i\}$, respectively, while $K^C_m(x_i)$, $K^C_m(y_i)$ and $K^C_m(z_i)$ denote their highest values (KLM). It is of interest to explore the KL and KLM complexities of the product of time series in dependence on the complexity of single ones.

## 3. Data and computations

### 3.1. Experimental details

Data we needed for the nonlinear dynamics in this paper we have obtained from the underground low-level laboratory in Belgrade. The special designed system for radon reduction, used in laboratory is consist of three stage: (a) The active area of the laboratory is completely lined up with aluminium foil of 1 mm thickness, which is hermetically sealed with a silicon sealant to minimize the diffusion of radon from surrounding soil and concrete used for construction, (b) the laboratory is continuously ventilated with fresh air, filtered through one rough filter for dust elimination followed by the battery of coarse and fine charcoal active filters and (c) the parameters of the ventilation system are adjusted to give an overpressure of about 2 mbar over the atmospheric pressure. The radon monitor is used to investigate the temporal variations in the radon concentrations. For this type of short-term measurements the SN1029 radon monitor was used (manufactured by the Sun Nuclear Corporation). The radon monitor device records radon and atmospheric parameters readings every 2 h in the underground laboratory. The data are stored in the internal memory of the device and then transferred to the personal computer. The data obtained from the radon monitor for the temporal variations of the radon concentrations over a long period of time enable the study of the short-term periodical variations (Udovičić et al., 2011). The distribution of frequencies of measured $^{222}$Rn concentration values is depicted in Fig. 1. It can be seen that the peak of distribution is about 10 Bq m$^{-3}$. The presence on indoor radon depends on large number of factors and Maxwell-like distribution of frequencies of measured values indicates probabilistic character of radon appearance in some room. Some authors presented results of similar measurements as log-normal distribution (Bossew, 2010).

*3.2 Computation of complexity measures*

For complexity analysis, we have used three time series of indoor: (1) $^{222}$Rn concentration, (2) air temperature and (3) air humidity (Fig. 2). Using the calculation procedure outlined in subsections 2.2-2.3, we have computed (a) complexity spectra of $^{222}$Rn concentration (Rn); product of $^{222}$Rn concentration and indoor air temperature time series (RnxT); indoor air humidity (RnxH) and product of $^{222}$Rn concentration, indoor air temperature and air humidity time series (RnxTxH) and (b) the KL and KLM for these series. Before computation procedure the time series were normalized on their highest values. The length of all time series used was $N=4173$. The computations are carried out for the period 1 January – 31 December 2009 using the LZA algorithm. Let us note that Hu et al. (2006) have showed that the complexity for a random sequence can be considerably larger than 1.

## 4. Results and disscusion

The concentration of $^{222}$Rn in some room is a result of dynamical steady-state of number of parameters. It can be expected that stronger as usual influence of some of them or absence of other one can result either in higher or lover values of $^{222}$Rn concentration in the monitored area. Thus, $^{222}$Rn concentration strongly depends on parameters of underground environment (Viñas et al., 2007). According to Udovičić et al. (2011) in the long term there exists a clear influence of indoor air temperature and relative humidity on $^{222}$Rn concentration. Further, in the same paper it is underlined that concerning the radon daughters, the relative humidity indoors contributes to the aerosol density and keeps the radon daughters in the indoor air. Although, in the last decade a vast number of experimental evidence has been offered about this issue (Choubey et al., 2011; Barbosa et al., 2010; Kamra et al., 2013), we still have no enough knowledge about insights of the influence of the indoor air parameters on the $^{222}$Rn concentration variability. One of the reasons for that is nonlinearity of these dependences, which can not be elaborated by the traditional mathematical methods (Seftelis et al., 2008). Thus, it seems that the complexity analysis offers more quantitative measures in explanation this phenomenon.

The results of complexity analysis are given in Fig. 3. This figure depicts the KL spectra of the following time series: (a) $^{222}$Rn concentration (Rn), (b) product of $^{222}$Rn concentration and indoor air temperature (RnxT), (c) indoor air humidity (RnxH) and (d) product of $^{222}$Rn concentration, indoor air temperature and indoor air humidity (RnxTxH), where all time series are normalized on their highest value. The peaks in spectra show the KLM. This parameter could be considered as a better indicator of the complexity comparing to the KL, which is not always suitable measure of the complexity. In particular, this is enhanced in the case of asymmetrical distributions (Nikolić-Đorić, personal communication). Looking at panels it is seen that for the RnxT sequence its KLM (Fig. 3b) is just slightly different comparing to the Rn one (Fig. 3b). Practically, there are no differences in their maximal Kolmogorov complexities. However, in the RnxH spectrum (Fig. 3c) and the RnxTxH one (Fig. 3d) the KLM values are lower then for the Rn spectrum. Note that a process that is least complex has a Kolmogorov complexity value near to zero, whereas a process with highest complexity will have KL close to one. This measure can be also considered as a measure of randomness. Thus, a value of the KL near zero is associated with a simple deterministic process like a periodic motion, whereas a value close to one is associated with a stochastic process (Ferreira, 2003). Accordingly the KLM values, which are large for the RnxT spectra (0.937), points out the presence of stochastic component in influence of indoor air temperature on $^{222}$Rn concentration. The other two calculated KLM complexities [RnxH (0.865) and RnxTxH (0.850) spectra] indicate that there exists a source of influence, which reduces the complexity of $^{222}$Rn concentration. To our opinion it could be attributed to (i) the fact that relative humidity indoors contributes to the aerosol density and keeps the radon daughters in the indoor air and (ii) nonlinearities in relation between $^{222}$Rn concentration and indoor air humidity (Nikolić-Đorić, personal communication). Finally, we have plotted the diagram KL complexity versus Sample Entropy (SE) to see behavior of time series. The KL measure has been often used for evaluation of the randomness present in time series, while entropy is also used to characterize the complexity of a time series. Thus, approximate entropy with a biased statistic, is effective for analyzing the complexity of noisy, medium-sized time series (Pincus, 1995). Richman and Moorman (2000) proposed another statistic, i.e. sample entropy (SE), which is unbiased and less dependent on data. It means that time series having the lower SE has less randomness components in their behavior. From Fig. 4 it is clearly see a strong correlation between these two measures indicating degree of influence of indoor air temperature and relative humidity on $^{222}$Rn concentration.

## 5. Concluding remarks

In this paper we have suggested a complexity measure based method for studying the dependence of $^{222}$Rn concentration time series on indoor air temperature and humidity. For that purpose we have used indoor $^{222}$Rn concentration time series measured during the 2009 in the Low-Background Laboratory for Nuclear Physics at the Institute of Physics in Belgrade. In that sense the following points can be enhanced.

(1) Since the method suggested is based on the Kolmogorov complexity measure (KL), which does not carries the whole information about the complexity of the time series, we have introduced: (i) the sequence of the KL, (ii) the Kolmogorov complexity highest value in the sequence (KLM) and (iii) the KL of the product of two time series.

(2) The complexity of time series can be lost due to reduction in functioning the system or process represented by that time series. It means that there exists a source, which causes that time series becomes rather uniform than random. In our case that is detection of indoor $^{222}$Rn concentration is under influence of some parameters. Which of these parameters contribute to reducing the complexity of detection indoor $^{222}$Rn concentration, we can establish through computing the complexity of the product of two or more time series obtained by measurements.

(3) We have calculated sequences of the following time series: (a) $^{222}$Rn concentration (Rn), (b) product of $^{222}$Rn concentration and indoor air temperature (RnxT), (c) indoor air humidity (RnxH) and (d) product of $^{222}$Rn concentration, indoor air temperature and indoor air humidity (RnxTxH), where all time series are normalized on their highest value. Also we have computed the KL and KLM of the aforementioned time series.

(4) The calculated KLM values, which are large for the RnxT sequence (0.937), points out the presence of stochastic component in influence of indoor air temperature on $^{222}$Rn concentration. The other two calculated KLM complexities [RnxH (0.865) and RnxTxH (0.850) sequences] indicate that there exists a source of influence, which reduces the complexity of $^{222}$Rn concentration. To our opinion it can be attributed to (i) the fact that relative humidity indoors contributes to the aerosol density and keeps the radon daughters in the indoor air and (ii) nonlinearities in relation between $^{222}$Rn concentration and indoor air humidity.

(5) We have plotted the diagram KL complexity versus sample entropy (whose lower values indicate on less randomness components in the time series). There exists a strong correlation between these two measures indicating degree of influence of indoor air temperature and relative humidity on $^{222}$Rn concentration.

**Acknowledgements**

This paper was realized as a part of the project "Studying climate change and its influence on the environment: impacts, adaptation and mitigation" (43007) financed by the Ministry of Education and Science of the Republic of Serbia within the framework of integrated and interdisciplinary research for the period 2011-2014 and as a part of the project "Biosensing Technologies and Global System for Long-Term Research and Integrated Management of Ecosystems" (43002).

**Figure captions**

Fig. 1. Distribution of frequencies of measured $^{222}$Rn concentrations.

Fig. 2. Time series for (a) $^{222}$Rn concentration, (b) indoor air temperature and (c) indoor air humidity, created from data obtained from the Low-Background Laboratory for Nuclear Physics at the Institute of Physics in Belgrade (Serbia) for the period 1 January – 31 December 2009.

Fig. 3. The Kolmogorov complexity (KL) sequences of the following time series: (a) $^{222}$Rn concentration (Rn), (b) product of $^{222}$Rn concentration and indoor air temperature (RnxT), (c) indoor air humidity (RnxH) and (d) product of $^{222}$Rn concentration, indoor air temperature and indoor air humidity (RnxTxH). All time series are normalized on their highest value.

Fig. 4. Kologorov complexity (KL) versus sample entropy (SE) for time series used in Fig. 5.

Figure 1.

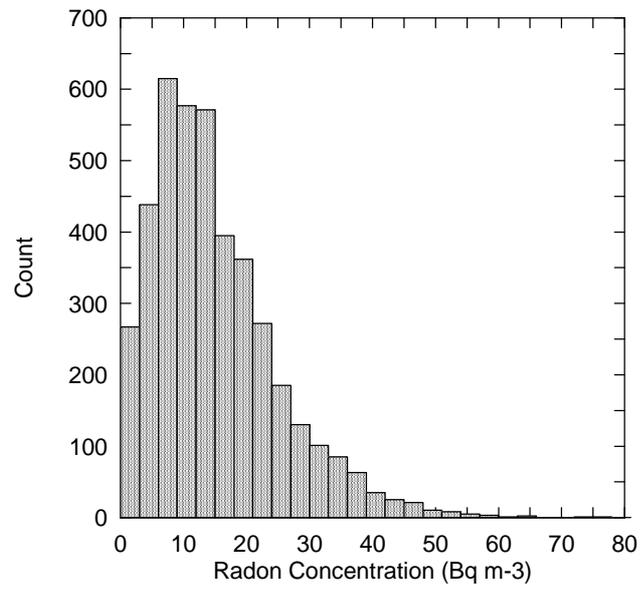

Figure 2.

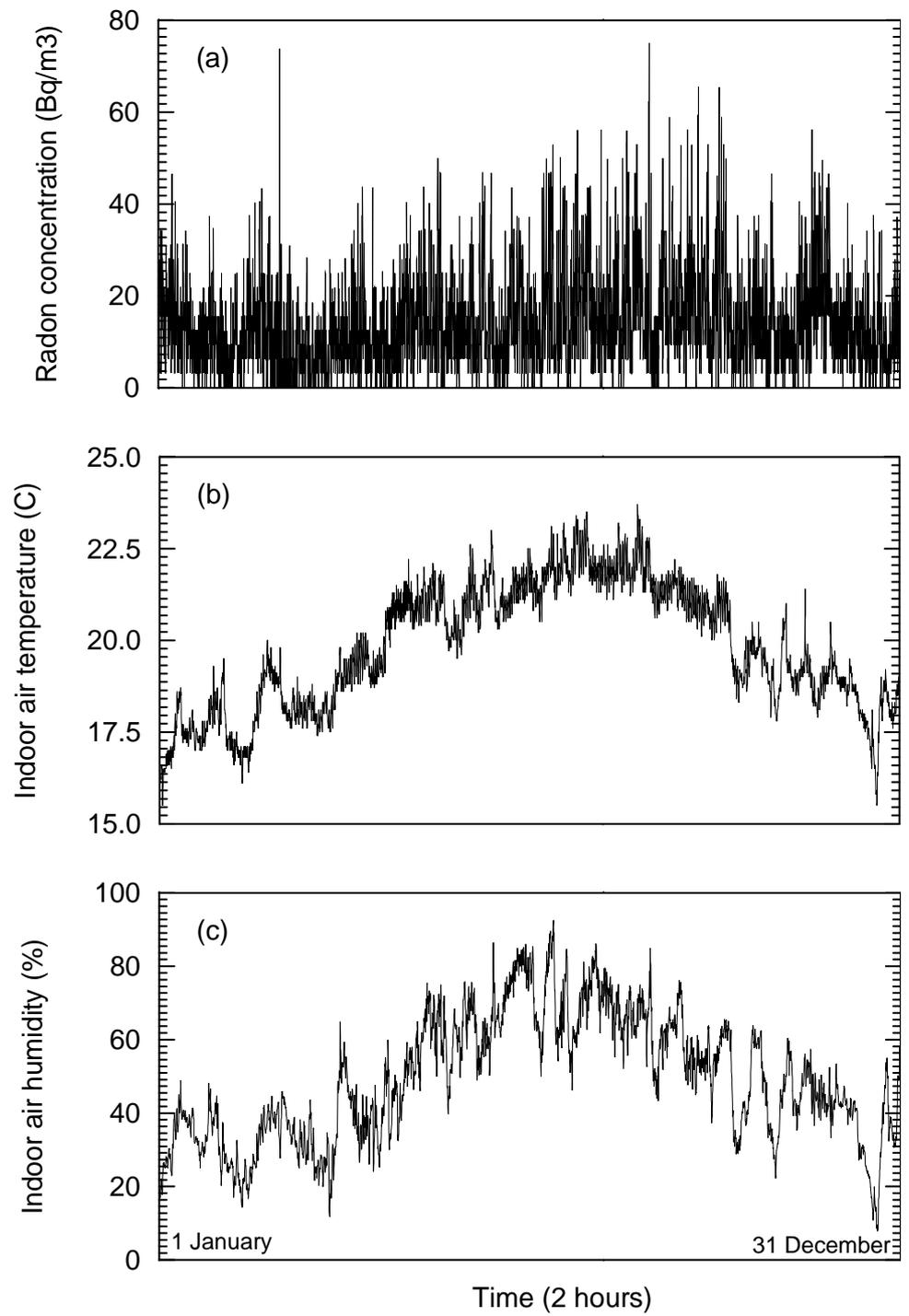

Figure 3.

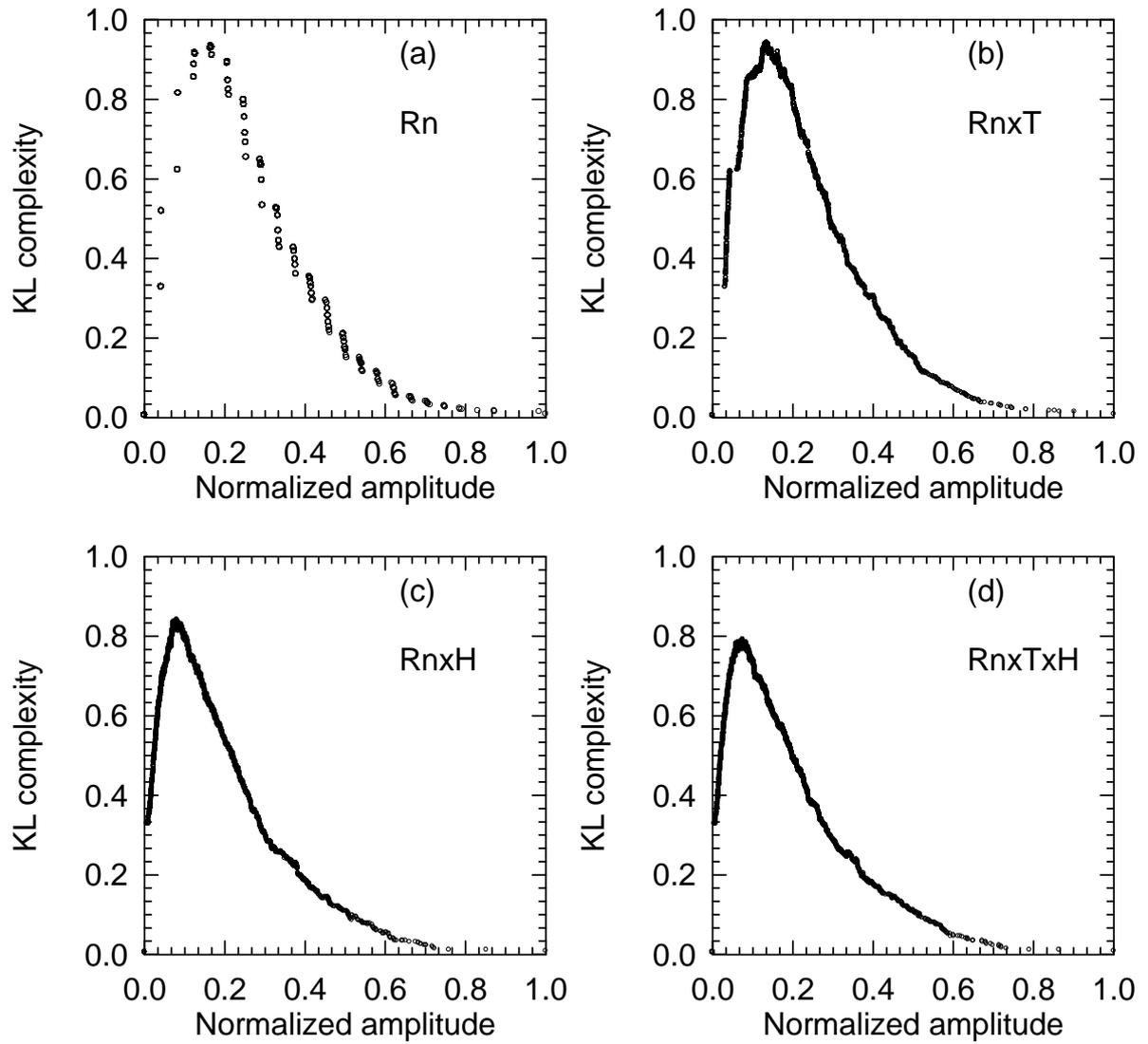

Figure 4.

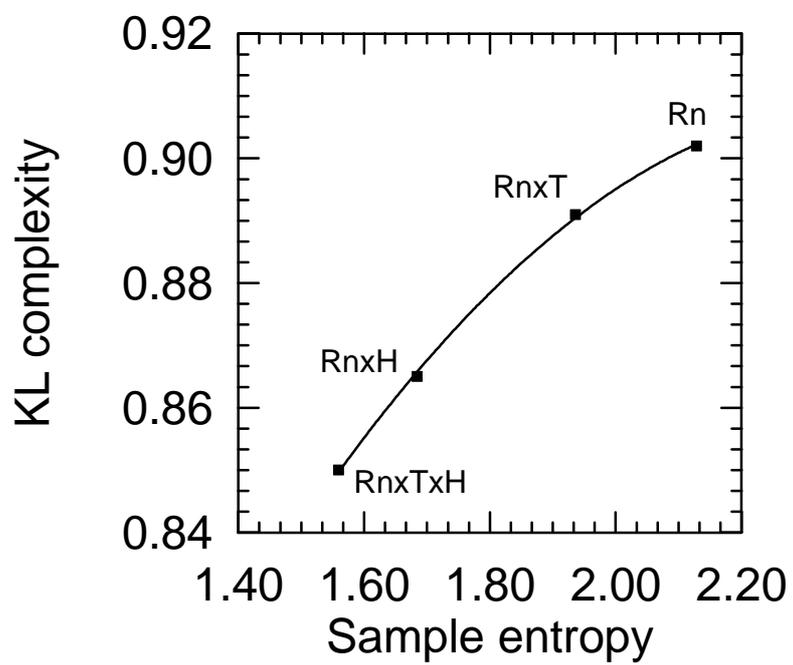